\documentclass[useAMS,usenatbib]{mn2e}
\usepackage{graphicx,amsmath,multicol}
\usepackage{psfrag}
\usepackage[normalem]{ulem}
\usepackage[dvipsnames]{xcolor}
\definecolor{webgreen}{rgb}{0,.5,0}
\definecolor{webbrown}{rgb}{.6,0,0}
\usepackage[pdfpagelabels]{hyperref}
\hypersetup{%
   colorlinks=true,hyperfootnotes=false,%
   breaklinks=true,%
   plainpages=false, bookmarksnumbered, bookmarksopen=true,
 bookmarksopenlevel=1,%
   urlcolor=webbrown, linkcolor=RoyalBlue, citecolor=webgreen,%
}

\def \be{\begin{equation}}
\def \ee{\end{equation}}
\def \bea{\begin{eqnarray}}
\def \eea{\end{eqnarray}}

\def \la {\langle}
\def \ra {\rangle}

\newcommand       \apj          {ApJ}
\newcommand       \apjl         {ApJL}
\newcommand       \apjs         {ApJS}

\newcommand       \mnras        {MNRAS}
\newcommand       \araa         {Ann. Rev. Astr. Astr.}
\def \msun {M_\odot}

\def\lesssim{\mathrel{\hbox{\rlap{\hbox{\lower4pt\hbox{$\sim$}}}\hbox{$<$}}}}
\def\gtrsim{\mathrel{\hbox{\rlap{\hbox{\lower4pt\hbox{$\sim$}}}\hbox{$>$}}}}

\title[Superbubble Feedback]{
In a hot bubble: why superbubble feedback works, but isolated supernovae do not?
}

\author[Prateek Sharma, Arpita Roy, Biman B. Nath, Yuri Shchekinov]
{Prateek Sharma$^1$ \thanks{prateek@physics.iisc.ernet.in},  Arpita Roy$^{1,2}$, Biman B. Nath$^2$, 
Yuri Shchekinov$^3$\\
$^1$Joint Astronomy Programme and Department of Physics, Indian Institute of Science, Bangalore 560012, India\\
$^2$Raman Research Institute, Sadashiva Nagar, Bangalore 560080, India\\
$^3$ Department of Physics, Southern Federal University, Rostov on Don 344090, Russia\\
}

\begin{document}

\maketitle

\label{firstpage}

\begin{abstract}

Using idealized 1-D Eulerian hydrodynamic simulations, we contrast the behavior of isolated supernovae 
with the superbubbles driven by multiple, collocated supernovae. Continuous energy injection via successive supernovae 
going off within the hot/dilute bubble maintains a strong termination shock. This strong shock keeps the superbubble over-pressured and drives 
the outer shock well after it becomes radiative.
Isolated supernovae, in contrast, with no further energy injection, become radiative quite early ($\lesssim 0.1$ Myr, 10s of pc), 
and stall at scales $\lesssim 100$ pc.
We show that isolated supernovae lose almost all of their mechanical energy by a Myr, but superbubbles can 
retain up to $\sim 40\%$ of the input energy in form of mechanical energy over the lifetime of the star cluster (few 10s 
of Myr). 
These conclusions hold even in the presence of realistic magnetic fields and thermal conduction.
We also compare various recipes for implementing supernova feedback in numerical simulations. For various feedback prescriptions we 
derive the spatial scale below which the energy needs to be deposited for it to couple to the interstellar 
medium (ISM). 
We show that a steady thermal wind within the superbubble appears only for a large number ($\gtrsim 10^4$) of supernovae. 
For smaller clusters we expect multiple internal shocks instead of a smooth, dense thermalized 
wind. 

\end{abstract}


\section{Introduction}

Gravity and dark energy govern the structure in the universe at the largest scales but complex baryonic 
processes like cooling, heating, self-gravity and  star formation are important at galactic scales (e.g., \citealt{spr05}). 
Numerical simulations have made tremendous progress in understanding galaxy formation, starting from pure gravitational 
N-body simulations to the current models which try to model the aforementioned complex processes. Modeling 
the gravitationally-interacting dark matter is straightforward in principle, and only limited by the available computing power. But the modeling 
of baryonic processes is rather involved. In particular, there is no consensus on which baryonic 
processes are important and how they should be implemented numerically. Given the dynamic range of 
scales, from large scale structure (10s of Mpc) to an individual star forming cloud ($\sim$pc), simulations have 
to resort to unresolved `subgrid' models for star formation and feedback due to star formation (e.g., \citealt{nav93,ger97,spr03,gue11,hop12} 
and references therein). While different star formation recipes seem to give similar star formation histories and
stellar mass distributions, provided molecular clouds are resolved (e.g., \citealt{hop11}), simulations are quite 
sensitive to the various feedback prescriptions (e.g., thermal feedback due to supernovae, momentum injection via dust 
absorbing/scattering photons produced by massive stars and supernovae) even with high resolution.

Stars form in clusters and super-star-clusters (100s to $10^6$ stars) of various sizes and in different environments, ranging from
 low density galactic outskirts to dense galactic centers (see \citealt{pro10} for a review). These clusters are observed to disrupt the dense 
 molecular clouds in which they are born (e.g., see \citealt{lei89,zha01}). This stellar feedback (due to strong radiation, stellar winds and 
 supernovae) 
disperses cold gas and suppresses further star formation. Because of the stellar initial mass function (IMF) and the main sequence lifetimes, the energy input is roughly constant 
per unit time over the life time of stars more massive than $8 \msun$ ($\sim 50$ Myr; \citealt{mcc87}). Therefore, 
stellar feedback is sometimes  modeled as a constant luminosity driven blast wave (\citealt{wea77,mac89,ger97}). A
superbubble (SB) expands faster than an isolated supernova remnant (SNR) because of continuous energy injection, 
and suffers smaller cooling losses because most supernovae (SNe) go off in a low density bubble.

The hot bubble breaks out through the gas disk if the outer shock driven by overlapping SNe crosses 
the scale height with a sufficient Mach number within the starburst lifetime
(e.g., see section 2 of \citealt{roy13}). After breakout the hot, metal-rich stellar ejecta is spread out into the galactic 
halo via the Rayleigh-Taylor instability. Spreading of metals over large scales is required to explain the high metallicity 
observed in the intergalactic medium (IGM), far away from the stellar disk (\citealt{tum11}).

Standard models for feedback through multiple SNe assume that a fraction $\gtrsim 0.3$ of the total explosion energy is retained in the hot ISM (e.g., \citealt{strickland09}). 
This fraction is much larger than the estimates for a single SNR ($\sim 0.1$) after the radiative phase at $\lesssim 0.1$ Myr (e.g., \citealt{cox72, che74,tho98}). Moreover, 
there are further radiative and adiabatic losses, such that over the timescale of order tens 
of Myr, the available mechanical energy in the bubble and shell is a negligible ($\ll 0.1$)  fraction of the initial supernova (SN) energy input.
 Recently \citet{nath13} argued that SBs created by multiple SNe in a star cluster are more effective as a feedback mechanism, in comparison
 to incoherent SNe. Recent simulations of interacting SNe by 
 \citealt{vas14} show that the fractional energy retained as thermal energy in the hot ISM can be as large as $\sim 0.1\hbox{--}0.3$ 
 only if the explosions are spatially and temporally correlated such that the radiative losses are effectively compensated by new explosions.
 SBs produced by compact star clusters are expected to satisfy this condition. In this paper we assume all SNe to
 be coincident, a good approximation if the SNR, when it becomes radiative, encompasses the whole cluster. This makes SBs a more 
 effective feedback agent which is responsible for magnificent galactic outflows.

In addition to elucidating differences between isolated SNe and SBs, we compare 
various methods of injecting  SN energy in numerical simulations of the ISM. Different feedback processes, 
such as radiation pressure, photoionization, and cosmic rays, are important in 
explaining outflows in galaxies of different masses and redshifts (e.g., \citealt{hop12} and references therein). 
In this paper we only focus on the most important feedback component, namely, thermal feedback due to SNe. 
We find that the SN thermal energy must be deposited over a sufficiently small volume for it to create a hot bubble  and to 
have an impact on the surroundings. For a large energy deposition radius the cooling time is shorter 
than the thermalization timescale and thermal feedback is artificially suppressed.
Most early implementations of SN feedback suffered from this problem (see \citealt{ger97} and 
references therein).

The formation of a low density bubble is essential for thermal SN feedback to work because 
the energy of subsequent SNe is deposited in the low density  bubble and is not radiated instantaneously; 
cooling is only restricted to the outer shock. While this fact has been appreciated (e.g., \citealt{gne98,jou06}), we  present 
{\em quantitative} conditions for the formation of a strong shock and a hot, dilute bubble for different thermal feedback 
prescriptions. The main culprit responsible for the inefficacy of thermal feedback, in both Eulerian (e.g., \citealt{tas06,dub08}) 
and Lagrangian (\citealt{spr03,sti06}) simulations, is the lack of resolution. In reality, a SN affects its surroundings starting at 
small ($\ll$ 1 pc) scales by launching very 
fast ejecta ($\sim 10^4$ km s$^{-1}$) into the ISM (e.g., see \citealt{che77}). Simulations in which SN energy is not
deposited at small enough scales have to resort to artificial measures (turning off cooling for several energy injection timescales, 
depositing energy into artificial `hot' phase that does not cool, etc.) in order for feedback to have any impact at all on the ISM 
(e.g., \citealt{ger97,spr03,dub08}). These measures are necessary for including the effects of supernova heating in large-scale simulations.

The continuous injection of mass and energy by SNe deep within the hot SB is expected to launch a steady wind 
(as first calculated by \citet{che85};  hereafter CC85). CC85 obtained analytic solutions for a superwind assuming a constant 
thermal energy and mass input rate with an injection radius. By modeling realistic SNe as fast moving ejecta within 
SBs we show that a steady CC85 wind is obtained only if a large number of SNe ($\gtrsim 10^4$) go off within 
the star cluster. For a smaller number of SNe, an individual SN's  kinetic energy is not thermalized within a small injection radius and 
there is an unsteady outflow. This should have implications for works that simply assume a CC85 wind within the SB.
    
Our paper presents analytic results that can be readily used for the numerical implementation of thermal feedback using various recipes.
These analytic criteria are verified and extended using idealized numerical simulations. The fundamental difference between isolated
supernovae and superbubbles -- that the former is ineffective on galactic scales -- is highlighted.

The paper is organized as follows. In section \ref{sec:fb_pres} we describe various ways of implementing thermal feedback due to 
SNe. Section \ref{sec:num_sim} describes the numerical setup used to study SNe and SBs.
 In section \ref{sec:anal} we present different analytic criteria for feedback to work with a range of feedback prescriptions. 
We also derive the conditions for obtaining a thermalized CC85 wind within a SB. In section \ref{sec:results}
we present 1-D numerical simulations of different feedback recipes with and without cooling, and compare the results with 
our analytic estimates. 
We also briefly discuss the effects of magnetic fields and thermal conduction. In section \ref{sec:conc} we discuss the implications of our 
results on galaxy formation, feedback simulations, and on galactic wind observations.
 
\section{ISM \& SN feedback prescriptions}
\label{sec:fb_pres}
 Although ISM is multiphase and extremely complex, for simplicity we consider a uniform, static model with a given 
 density (typically $n=1$ cm$^{-3}$) and temperature ($10^4$ K corresponding to the warm ISM). We do not consider 
 stratification because the disk scale height is typically a few 100 pc and the fizzling 
of SN feedback is essentially a small scale problem. Moreover, the scales of interest (100s of pc; few Myr) are much 
bigger than the cluster size and the local ISM/circumstellar inhomogeneities.
 For simplicity we also assume that all SNe 
 explode at $r=0$. This is a good approximation because the size of a typical (super) star-cluster is smaller than the bubble size at the 
 beginning of the radiative phase. 
 
The SN prescriptions that we consider in our analytic estimates and numerical simulations cover the full range of methods used in the literature.
These are:
 
\begin{enumerate}
 
 \item {\em Kinetic Explosion Models} (KE): In these models the SN energy ($E_{\rm ej}$, chosen to be 1 Bethe $\equiv 10^{51}$ erg)
  is given to a specified ejecta mass ($M_{\rm ej}$, chosen to be 1 $\msun$) distributed uniformly with an ejecta density $\rho_{\rm ej} 
  = 3 M_{\rm ej}/[4 \pi r_{\rm ej}^3$] within an ejecta radius $r_{\rm ej}$. The ejecta velocity is homologous with $v_{\rm ej}(r) = v_0 (r/r_{\rm ej})$ within the ejecta; the 
  normalization is such that the kinetic energy of the ejecta is $E_{\rm ej}$; i.e., $v_0=(10E_{\rm ej}/3M_{\rm ej})^{1/2}$. The ejecta 
  temperature is taken to be small ($T_{\rm ej}=10^4$ K). 
   After every (fixed) SN injection time $t_{\rm SN}$
   the innermost $r_{\rm ej}$ of the volume is {\em overwritten} by the ejecta density and velocity, thereby pumping SN energy into 
   the ISM. After the reverse shock propagates toward the bubble center, once the swept-up mass is comparable to the ejecta mass, 
   the bubble density structure is fairly insensitive to the ejecta density distribution (\citealt{tru99}). 
   This model most closely resembles a physical SNR in early stages at small ($\ll$ 1 pc) scales when 
   the SN `piston' at large speed rams into the ISM. 
   This prescription is not widely used in galaxy formation simulations (there are some exceptions, e.g., \citealt{tan05}).
 
 \item {\em Thermal Explosion Models} (TE): In these models the energy is deposited within the ejecta radius in form of thermal energy 
 at an interval of $t_{\rm SN}$. There are two variants of this model. In one class, the mass and internal energy densities are {\em overwritten} 
 within the ejecta radius ($r_{\rm ej}$) such that the {\em uniformly distributed} ejecta thermal energy is $E_{\rm ej}$ (1 Bethe) and the 
 {\em uniformly distributed} 
 ejecta mass is $M_{\rm ej}~(1 \msun)$. We abbreviate these models as TEo and they behave like KE models. The second class of models is 
 where we {\em add} (in contrast to overwrite) the ejecta mass (with uniform density) to the preexisting mass and the ejecta 
thermal energy (distributed uniformly) to the preexisting internal energy within $r_{\rm ej}$. We refer to these models as TEa. 
There are significant differences between the TEa and TEo/KE models in presence of cooling because TEa ejecta can become radiative 
if thermal energy is added to a dense ISM. Most models in the literature are analogous to TEa models with some variations 
(e.g., \citealt{kat92}, \citealt{jou06}, \citealt{cre13}; some works such as \citealt{sti06} ,\citealt{tha00}, \citealt{age11} 
unphysically turn off cooling for some time for SN feedback to have an impact). Sometimes all the SN energy is deposited in a 
single grid cell (e.g., \citealt{tas06}) or in a single particle (e.g., SN particle method in section 3.2.4 of \citealt{ger97}).
 
 \item {\em Luminosity Driven Models} (LD): As discussed earlier, for typical IMFs, the mechanical energy input due to OB stars per unit time
  is roughly constant. This motivates a model in which internal energy and mass within an injection radius (denoted by $r_{\rm ej}$) 
  increase at a constant rate corresponding to internal energy $E_{\rm ej}$ and mass $M_{\rm ej}$ for each SN. Some 
  of the works that use this prescription are \citealt{che85,mac89,suc94,mac99,str00,coo08,roy13,rec13,pal13}. 

 \end{enumerate}
  
All our models are identified 
with the number of OB stars $N_{\rm OB}$  (which equals the number of SNe in the star cluster) or luminosity 
$L_{\rm ej} = E_{\rm ej}/t_{\rm SN} = N_{\rm OB} E_{\rm ej}/t_{\rm OB}$, where $t_{\rm OB}$ is the lifetime of the OB association 
(taken to be 30 Myr) and $t_{\rm SN}=t_{\rm OB}/N_{\rm OB}$ is the time interval between SNe. We note that
the overwrite models do not strictly conserve mass and energy. While the supernova energy is much larger than the overwritten 
thermal energy, one needs to choose a small enough ejecta radius such that the overwritten mass is subdominant relative to 
the ejecta mass. The explosion models are substantially slower compared to the smooth luminosity driven models because of very high 
temperatures created due to sudden energy injection in the former.  
 
\section{Numerical setup}
\label{sec:num_sim}
In this section we describe the numerical setup corresponding to our one-dimensional simulations discussed in section \ref{sec:results}.
Our numerical simulations use the ISM setup and the various feedback prescriptions described in section \ref{sec:fb_pres}. 
The ISM density and temperature are chosen to 1 cm$^{-3}$ and $10^4$ K, respectively, unless specified otherwise. The mean 
mass per particle is $\mu=0.62$ and per electron is $\mu_e=1.17$. The initial density and temperature are uniform, and the
velocity is zero. 

We use the grid-based {\tt ZEUS-MP} code in spherical, one-dimensional geometry (\citealt{hay06}) to solve the standard Euler 
equations with source terms mimicking SN energy/momentum/mass injection for the chosen feedback model,
and a sink term in the internal energy equation representing 
radiative cooling. Our equations are similar but not identical to Eqs. 20-22 in \citealt{roy13}.\footnote{The difference from \citet{roy13} is 
that the mass and energy source terms are not uniform in time. Moreover, different feedback heating prescriptions use different source 
terms; e.g., KE models use a momentum source term rather than a source term in the internal energy equation.} 
We note that the {\tt ZEUS} code does not conserve energy to the machine precision. However, we have confirmed that energy conservation
holds to better than 90\% in most of our runs. In fact, the Sedov-Taylor blast wave in 1-D spherical coordinates is one of the standard test problems presented in \citet{hay06}; the numerical results match analytic solutions very closely.
While most of our runs are hydro, in section \ref{sec:B_cond} we briefly discuss runs with simple 
models of magnetic fields and thermal conduction; various parameters for these runs are discussed there.
%
%

The radial  velocity is set to zero at the inner radial boundary and other fluid variables are copied in the ghost cells. Outflow boundary conditions
are applied at the outer radial boundary. Note that the outer boundary is out of causal contact. 
Since the setup produces strong shocks, we use the standard {\tt ZEUS} artificial viscosity to prevent unphysical oscillations at 
shocks (\citealt{sto92}). The CFL number of 0.2 is found to be more robust compared to the standard value of 0.5, and is used in 
all the simulations. 

The cooling function that we use is the collisional-ionization-equilibrium based solar metallicity table of \citet{sut93}, with the cooling function 
set to zero below $10^4$ K.  The cooling step is implemented via operator splitting using the 
semi-implicit method of \citet{sha10}. Cooling is subcycled and the number of subcycles is limited to be less than 100.

Most of our runs use a 1024 resolution grid extending from 1 pc to 2 kpc. A logarithmically spaced grid is used to better resolve 
smaller radii; there are equal number of grid points covering 1 pc to $\sqrt{2000}$ pc and $\sqrt{2000}$ pc to 2 kpc. Some runs with stronger SN feedback use a larger spatial extent (c.f. $N_{\rm OB}=10^6$ runs in Fig. 
\ref{fig:rho_vs_r_CC85}), and some uniform very high resolution runs (16384 grid points; c.f. Fig. \ref{fig:d_T}) use a smaller 
extent (1 to 200 pc). All our simulations (except the very high resolution ones that are run for 3 Myr) are run for 30 Myr, typical age of a young star cluster.
  
 \section{Analytic criteria}
 \label{sec:anal}
 
 In this section we present the analytic criteria that need to be satisfied for various feedback models discussed in section 
 \ref{sec:fb_pres} to work. These analytic estimates help us understand the results of numerical simulations discussed in section 
 \ref{sec:results}. While radiative cooling is the most discussed phenomenon in the context of fizzling SN 
 feedback, the feedback prescription should satisfy additional constraints for the energy input to couple realistically
  with the ISM. A recurring concept in what follows is that of thermalization; i.e., in order to be effective the input energy 
should have time to couple to the ISM before it is radiated or is overwritten. In section \ref{sec:CC85} we show that a
steady superwind within a superbubble, as envisaged by \citet{che85}, occurs only if the number of SNe is sufficiently 
large. 
 
 In the following sections we derive upper limits on the ejecta radius within which the feedback energy must be deposited for it to be 
 effective. We can easily convert this radius limit into a critical mass resolution needed 
 in smooth-particle-hydrodynamics (SPH) simulations; namely, $n_{\rm nbr}  m_{\rm crit} \approx (4\pi/3) \rho r_{\rm crit}^3$, where $\rho$ is 
 the ISM density, $n_{\rm nbr}$ is the number of neighbors used in the SPH smoothing kernel and $m_{\rm crit}$ is the maximum SPH gas 
 particle mass required for feedback to work. 
 
 \subsection{Energy coupling without cooling}
 \label{sec:nocool}
 \subsubsection{Ejecta radius constraint for overwrite models}
 \label{sec:over}
 In models where energy within the ejecta radius is overwritten (KE, TEo) the ejecta radius should be smaller than 
 a critical radius ($r_{\rm ej} \lesssim r_{\rm crit}$) for the input energy to get coupled to the ISM.
 The critical radius equals the Sedov-Taylor shock radius at $t_{\rm SN}$,\footnote{We use the Sedov-Taylor expression for the bubble radius in 
 Eq. \ref{eq:r_therm} because the shock quickly transitions from a free-expanding to a Sedov-Taylor state; the 
 Sedov-Taylor radius (when the swept up ISM mass equals the ejecta mass) is $r_{\rm ST} \equiv (3M_{\rm ej}/4\pi\rho)^{1/3} 
 \approx 2.5 ~{\rm pc}~M_{{\rm ej},\odot}^{1/3} n^{-1/3}$, much smaller than the estimate in Eq. \ref{eq:r_therm}, where $M_{{\rm ej},\odot}$ 
 is the ejecta mass in solar units.} the time lag between SNe,
 \be
 \label{eq:r_therm}
 r_{\rm crit} \equiv  \left ( \frac{ E_{\rm ej}t_{\rm SN}^2}{\rho} \right) ^{1/5} \approx 50~{\rm pc}~n^{-1/5} E_{{\rm ej},51}^{1/5} t_{{\rm SN},0.3}^{2/5},
 \ee
 where $\rho$ ($n$) is the ISM (number) density (assuming $\mu=0.62$), $E_{{\rm ej},51}$ is the ejecta 
 energy in units of $10^{51}$ erg, and $t_{{\rm SN},0.3}$ is the time between consecutive SNe in units of $0.3$ Myr. 
 If the ejecta radius is larger than this value the ejecta energy is overwritten before it can push the outer shock. Thus,
 in such a case, the input SN energy is overwritten without much affecting the ISM. 
 
 \subsubsection{Sonic constraint}
For thermal SN feedback to launch a strong shock the energy should be deposited over a small enough volume, 
such that the post-shock pressure is much larger than the ISM pressure. This is equivalent to demanding
the outer shock velocity to be much larger than the sound speed in the ISM. The shock velocity ($v_{\rm OS} \equiv dr_{\rm OS}/dt$; $r_{\rm OS}$ is the outer shock radius), expressed in terms of
the shock radius in Sedov-Taylor stage, is $v_{\rm OS} \approx 0.4 E_{\rm ej}^{1/2} \rho^{-1/2} r_{\rm OS}^{-3/2}$ for an isolated 
SN and $v_{\rm OS} \approx 0.6 L_{\rm ej}^{1/3} \rho^{-1/3} r_{\rm OS}^{-2/3}$ for a luminosity driven SB (\citealt{wea77}).   
The condition for a strong shock for an isolated SN is ($v_{\rm OS} \lesssim a_T$; $a_T$ is the ISM isothermal sound speed)
\be
\label{eq:r_SN}
r_{\rm ej} \lesssim 174~{\rm pc}~E_{\rm ej,51}^{1/3} n^{-1/3} T_4^{-1/3}
\ee
and for a SB is (see Eq. 3 in \citealt{sil09})
 \be
 \label{eq:r_SB}
 r_{\rm ej} \lesssim 1.5~ {\rm kpc}~L_{\rm ej,38}^{1/2} n^{-1/2} T_4^{-3/4},
 \ee
where $L_{\rm ej, 38}$ is the ejecta luminosity ($L_{\rm ej}=E_{\rm ej}/t_{\rm SN}$ for explosion models) in units 
of $10^{38}$ erg s$^{-1}$  (this corresponds to $N_{\rm OB}=100$ over $t_{\rm OB}=30$ Myr) and $T_4$ is the ISM temperature in 
units of $10^4$ K. 
 
The sonic constraint ($v_{\rm OS} \lesssim a_T$) is typically less restrictive than the compactness requirements due to cooling in 
a dense ISM (see next section). \citet{tan05}, who considered supernova feedback in the hot ISM ($\sim 10^7$ K) of galaxy clusters 
and elliptical galaxies, found that the shock can quickly (when outer radius is only $\approx 20$ pc; see Eq. \ref{eq:r_SN}) decelerate to attain the sound speed in the hot ISM. After this the outer shock propagates as a sound wave.
While the sound wave can spread the SN energy over a larger radius ($\propto t$ for a sound wave, unlike a strong blast wave in which 
$r_{\rm OS} \propto t^{2/5}$),  energy dissipation is not as efficient as in shocks.
 
 \subsection{Energy coupling with cooling}
\label{sec:cool_cond}
 \subsubsection{Luminosity driven model}
 In luminosity driven models (LD), SN feedback does not fizzle out (in fact, the shock can get started) only if the cooling rate 
 is smaller than the energy deposition rate, i.e., $3L_{\rm ej}/4\pi r_{\rm ej}^3 \gtrsim n^2 \Lambda$ ($\Lambda[T]$ is the cooling 
 function), or
 \be
 \label{eq:lum_fizzle}
 r_{\rm ej} \lesssim 20~{\rm pc}~L_{{\rm ej},38}^{1/3} n^{-2/3} \Lambda_{-22}^{-1/3},
 \ee
 where $\Lambda_{-22}$ is the cooling function in units of $10^{-22}$ erg cm$^3$ s$^{-1}$.
 
 \subsubsection{Thermal explosion addition model}
 
 The above criterion (Eq. \ref{eq:lum_fizzle}) for the luminosity driven models is quite different from the criterion that we now derive for the widely 
 used thermal explosion models with energy and mass addition (TEa model in section \ref{sec:fb_pres}). Since energy is 
 {\em added} to the (possibly dense) pre-existing medium, cooling in this model can be substantial. In contrast, since the ejecta 
 density is low, cooling losses are smaller in the overwriting models (KE, TEo). For TEa models to launch a shock, radiative losses 
 over the timescale in which the shock from a point explosion reaches the ejecta radius,
 \be
 \label{eq:t_therm}
 t_{\rm ej} = E_{\rm ej}^{-1/2} r_{\rm ej}^{5/2} \rho^{1/2},
 \ee
 should be smaller than the energy deposited (here $\rho$ is the density of the medium in which energy is injected, not necessarily the ISM density); i.e., 
 \be
 \label{eq:cool_cond}
 n^2 \Lambda t_{\rm ej} \lesssim 3 E_{\rm ej}/4\pi r_{\rm ej}^3.
 \ee 
 Plugging in the expression for $t_{\rm ej}$, we get
 \be
 \label{eq:eadd_fizzle}
 r_{\rm ej} \lesssim 31~{\rm pc}~E_{\rm ej, 51}^{3/11} \Lambda_{-22}^{-2/11} n^{-5/11}.
 \ee
 This condition is much more restrictive than the one obtained by replacing $t_{\rm ej}$  in Eq. \ref{eq:cool_cond}
 by the CFL stability  timestep. Moreover, this is the appropriate timescale to use because the relevant timescale for the injected 
 energy to couple to the ISM is the thermalization time ($t_{\rm ej}$). 
  
 \citet{cre11} and \citet{dal12} have used similar arguments and derived results not too different from ours for the thermal explosion addition 
  models. A slight difference from our work is that they consider energy deposition over a resolution element (a necessity because of a larger range of 
  scales in cosmological galaxy simulations), but we allow for energy deposition
  over a resolved region. \citet{cre11} have expressed their resolution limit in terms of the cooling rate per unit mass and \citet{dal12} in terms of 
  the post-shock temperature; we use the cooling function ($\Lambda$) to express the critical radius within which the energy needs to be deposited.
  
\subsubsection{Overwrite models} 

 In models where the energy and mass densities are overwritten within $r_{\rm ej}$, the condition for overcoming cooling 
 losses and launching a shock is
 $$
 n_{\rm ej}^2 \Lambda  t_{\rm ej} \lesssim \frac{3 E_{\rm ej}}{4 \pi r_{\rm ej}^3},
 $$
 where ejecta number density $n_{\rm ej} = \rho_{\rm ej}/\mu m_p$ and $\rho_{\rm ej} = 3M_{\rm ej}/4\pi r_{\rm ej}^3$; note that 
 this expression is different from Eq. \ref{eq:cool_cond} in that the ejecta density is used instead of the ISM density.
 The overwrite models (KE, TEo) behave quite differently from addition (TEa, LD) models because a larger ejecta radius means a smaller density ejecta to which energy is added. Replacing 
the ISM density by the ejecta density in Eq. \ref{eq:t_therm} gives $t_{\rm ej} = E_{\rm ej}^{-1/2} r_{\rm ej}^{5/2} \rho_{\rm ej}^{1/2}$, 
and the condition for energy thermalization is
 \be
 r_{\rm ej} \gtrsim 0.003~{\rm pc}~M_{\rm ej,\odot}^{5/4} \Lambda_{-22}^{1/2} E_{\rm ej,51}^{-3/4}.
 \ee
 In order to avoid 
 radiative losses the ejecta radius should be {\em larger} than above. 
 This early cooling of the mass loaded SN ejecta, responsible for creating cold filaments in young SNe (e.g., \citealt{che95}),
 is physical (unlike fizzling out of the energy addition models) and should reduce the energy available to drive the 
SN. All our simulations use an ejecta radius much greater than this limit.
  
 \subsection{Conditions for CC85 wind}
 \label{sec:CC85}
 \citealt{che85} (hereafter CC85) found analytic solutions for a luminosity driven wind with a fixed injection radius. 
Luminosity injection is expected to drive both an outer shock bounding the bubble and a wind that shocks within the hot 
bubble at the termination shock (see Fig. 1 in \citealt{wea77}; see also LD run in Fig. \ref{fig:rho_vs_r_CC85}). In this section we show that for a small number of SNe (c.f. Eq. \ref{eq:CC_crit}) 
the  SN ejecta does 
not thermalize within the termination shock. In that case, the density inside the bubble is much lower than the CC85 wind because most SNe occur in 
the dilute bubble created by earlier SNe and the thermalization radius is comparable to the outer shock radius.
This has important implications on cooling and luminosity of SN ejecta. 
 
Following \citet{wea77}, the outer shock radius of a luminosity driven bubble is given by $r_{\rm OS} \approx (L_{\rm ej} t^3/\rho)^{1/5}$, velocity 
by $v_{\rm OS} \approx 0.6 r_{\rm OS}/t \propto t^{-{2/5}}$, and the post-shock pressure 
by $p_{OS} \approx 0.75 \rho v_{\rm OS}^2 \approx 0.27 L_{\rm ej}^{2/5} \rho^{3/5} t^{-4/5} $. Assuming a steady superwind, the 
ram pressure at the termination shock ($r_{\rm TS}$; the wind is assumed to be supersonic at this radius) is 
$\rho_{\rm TS} v_{\rm TS}^2 = \dot{M}_{\rm ej} v_{\rm TS}/(4 \pi r_{\rm TS}^2)$,  
where $v_{\rm TS}=(2 L_{\rm ej}/\dot{M}_{\rm ej})^{1/2}$ is the wind velocity, $\rho_{\rm TS}$ is the density upstream of the 
termination shock, and $\dot{M}_{\rm ej}$ is the mass injection rate. 
The wind termination shock ($r_{\rm TS}$) is located where the wind ram pressure balances the 
bubble pressure; i.e.,
$$
\frac{\dot{M}_{\rm ej} v_{\rm TS}}{4 \pi r_{\rm TS}^2} \approx 0.75 \rho v_{\rm OS}^2.
$$
Using $v_{\rm OS} \approx 0.6 L_{\rm ej}^{1/3} \rho^{-1/3} r_{\rm OS}^{-2/3}$ and $\dot M_{\rm ej}=2L_{\rm ej}/v_{\rm TS}^2$
gives,
\be
\label{eq:r_TS}
\frac{r_{\rm TS}}{r_{\rm OS}} \approx \left ( \frac{v_{\rm OS}}{v_{\rm TS}} \right )^{1/2} \approx 0.08 E_{\rm ej, 51}^{-1/12} M_{\rm ej, \odot}^{1/4} n^{-1/6} r_{\rm OS, 2}^{-1/3} t_{\rm SN, 0.3}^{-1/6},
\ee
 where $r_{\rm OS,2}$ is the outer shock radius in units of 100 pc, $t_{\rm SN,0.3}$ is the time between SNe normalized to 0.3 Myr (corresponding
 to $N_{\rm OB}=100$); we have used $L_{\rm ej}=E_{\rm ej}/t_{\rm SN}$ and $\dot{M}_{\rm ej}=M_{\rm ej}/t_{\rm SN}$. The ratio $r_{\rm TS}/r_{\rm OS}$ depends very weakly on time ($\propto t^{-1/5}$); this comes from the time dependence 
 of $r_{\rm OS}$ in Eq. \ref{eq:r_TS}. The reverse shock for an isolated SN very quickly (at the beginning of Sedov-Taylor stage) collapses to a point but the termination shock for a 
 SB is present at
 all times. Thus the non-radiative termination shock can power a SB long after the outer shock becomes radiative, unlike a SN which dies off shortly
 after the outer shock becomes radiative (see section \ref{sec:energy_budget} for our results from simulations).

The condition for the existence of a smooth CC85 wind is that the 
ejecta thermalization radius should be smaller than the termination shock radius. The superwind is mass loaded by previous 
SNe (the bubble density in the absence of mass loading is quite small because most of the mass is swept up in the outer shell). 
The swept up mass till radius $r$ in a CC85 wind is  
 $$
 M_{\rm sw} = \int_0^r 4 \pi r^{\prime 2} \rho_w (r^\prime) dr^\prime \approx \frac{\dot{M}_{\rm ej} r}{v_{\rm TS}} = \frac{M_{\rm ej} r}{t_{\rm SN} v_{\rm TS}},
 $$
 where $\rho_w(r^\prime)$ is the wind density profile; here we have assumed that the swept up mass is dominated by the supersonic portion of the wind.
 Now the thermalization radius (the radius within which the deposited energy is thermalized and which should correspond
 to CC85's injection radius) of the ejecta is where the swept up mass roughly equals the ejecta mass, or
 \be
\label{eq:rth_CC}
 r_{\rm th} \approx v_{\rm TS} t_{\rm SN} \approx 3~{\rm kpc}~E_{\rm ej, 51}^{1/2} M_{\rm ej, \odot}^{-1/2} t_{\rm SN, 0.3}.
\ee
 Since the thermalization radius is quite large, a thermalized CC85 solution will only occur for large clusters (with shorter 
 $t_{\rm SN}$, the time lag between SNe); 
for modest star clusters the ejecta will only thermalize beyond the termination shock. Of course, the thermalization radius cannot be smaller than the size of the star cluster launching the energetic ejecta. Using Eq. \ref{eq:r_TS} and $r_{\rm OS} \approx (L_{\rm ej} t^3/\rho)^{1/5}$, the termination shock radius can be expressed as
$$
r_{\rm TS} \approx 5~{\rm pc}~E_{\rm ej, 51}^{1/20} M_{\rm ej,\odot}^{1/4} n^{-3/10} t_{\rm SN,0.3}^{-3/10} t_{0.3}^{2/5},
$$
where $t_{0.3}$ is time in units of 0.3 Myr.
A CC85 solution will appear only if this termination shock radius is larger than the thermalization radius (Eq. \ref{eq:rth_CC}); i.e., if
\be
\label{eq:CC_crit}
t_{\rm SN,0.3} \lesssim 0.007 E_{\rm ej,51}^{-9/26} t_{0.3}^{4/13} n^{-3/13} M_{\rm ej,\odot}^{15/26}.
\ee
This means that $N_{\rm OB} \gtrsim 3500$ (recall that $t_{\rm SN}=t_{\rm OB}/N_{\rm OB}$,
 where $t_{\rm OB}=30$ Myr is the cluster lifetime and $N_{\rm OB}$ is the number of SNe) is required for a CC85 
 wind to appear by 30 Myr. Thus, a thermally driven CC85 wind occurs only for a sufficiently big starburst, 
 with a large mass loading, and at late times.
 
\section{Simulation  Results}
 \label{sec:results}
In this section we present the results from our one-dimensional numerical simulations. We vary the ISM density and SN 
injection parameters to assess when SN energy can significantly affect the ISM, both with and  without cooling. We also 
numerically verify the various analytic constraints presented in section \ref{sec:anal}. We discuss the structure of a radiative 
SB and compare the energetics of isolated SNe and SBs. While isolated SNe lose most of their mechanical energy by a few Myr, 
SBs can retain up to $\sim 40\%$ of the input energy long after the outer shock becomes radiative. Thus, SBs, and not isolated SNe, 
are the viable energy sources for global, galactic-scale feedback. In section \ref{sec:B_cond} we briefly discuss the impact of 
magnetic fields and thermal conduction on SBs.

 \subsection{Realistic SN shock (KE models)}
 
The SN shock is launched once a protoneutron star forms at the center of a massive evolved star (with size $\sim 10^{14}$ cm).  
In the ejecta dominated state (when the swept up ISM mass is less than the ejecta mass) the cold ejecta is dominated by kinetic 
energy (e.g., \citealt{tru99}).  In our `realistic' simulations (KE models; see section 2) we choose 
the ejecta to have a constant density and a velocity 
proportional to the radius (homologous expansion; this is a similarity solution for the freely expanding ejecta) within the ejecta. 
The SN shock develops a reverse shock after sweeping up its own mass in the ISM; this slows down the ejecta and communicates 
the presence of the ISM to the supersonic ejecta. 
In this section we compare the evolution of {\em adiabatic} (cooling is turned off) KE models with different parameters, highlighting the
 importance of having a small ejecta radius ($r_{\rm ej}$) even in absence of cooling for overwrite (KE, TEo) models. We have verified 
 that kinetic explosion (KE) and thermal explosion overwrite (TEo) models behave in a similar fashion.

\begin{figure}
\begin{center}
\psfrag{A}[cc][][1.][0]{$r({\rm pc})/N_{\rm OB,100}^{1/5}$}
\includegraphics[scale=0.5]{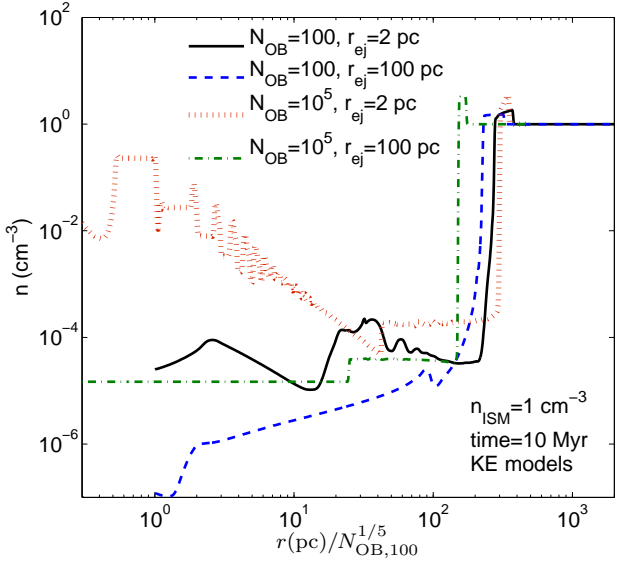}
\caption{ Number density as a function of radius (scaled to the self-similar scaling) for different parameters of  
realistic KE runs at 10 Myr. The outer shock is closer in for models using a larger ejecta radius 
because energy is overwritten before it can couple to the ISM. 
\label{fig:rho_vs_r}
}
\end{center}
\end{figure}
 
 Figure \ref{fig:rho_vs_r} shows the density profile as a function of radius (normalized to the self similar scaling, $r_{\rm OS} \approx [L_{\rm ej}t^3/\rho]^{1/5}$, where $L_{\rm ej}=E_{\rm ej}/t_{\rm SN}=E_{\rm ej} N_{\rm OB}/t_{\rm OB}$) for different realistic runs
 (results are similar for TEo models) 
 with $N_{\rm OB}=100,~10^5$ at 10 Myr. 
 The runs with a large ejecta radius (100 pc) give a smaller outer shock radius because most of the energy is overwritten without being 
 thermalized (see section \ref{sec:over} for a discussion). The problem is worse for larger $N_{\rm OB}$ (shorter $t_{\rm SN}$), as expected from
 Eq. \ref{eq:r_therm}. The normalized location of the outer shock falls almost on top of each other for a 
 small ejecta radius ($r_{\rm ej}=2$ pc). As expected, the shock is weaker, broader, and with a modest density jump for a smaller number of 
 SNe.
  
 \subsection{Comparison of adiabatic models}

While the KE model is most realistic, we expect other models in section \ref{sec:fb_pres} to give a similar location for the outer shock 
after the swept-up ISM mass equals the ejecta mass and the shock is in the Sedov-Taylor regime. The structure within the bubble depends on SN 
prescription, as we show in section \ref{sec:CC85_sims}.

\begin{figure}
\begin{center}
\psfrag{C}[][][1.][0]{$\propto t^{2/5}$}
\psfrag{D}[][][1.][0]{$\propto t^{3/5}$}
\includegraphics[scale=0.5,angle=0]{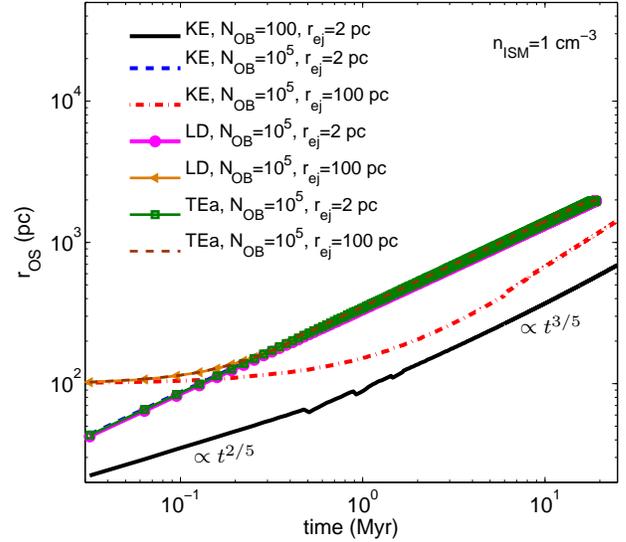}
\caption{The outer shock radius as a function of time for various runs using kinetic explosion (KE), luminosity driven (LD) and 
thermal explosion addition (TEa) models. The KE models give correct results only if the ejecta radius ($r_{\rm ej}$) is sufficiently 
small; otherwise energy is overwritten before getting coupled to the ISM. There is no such problem for energy addition and 
luminosity driven models. At early times
the outer shock radius scales with the Sedov-Taylor scaling ($r_{\rm OS} \propto t^{2/5}$) and later on, after many SNe go off, 
it steepens ($r_{\rm OS} \propto t^{3/5}$). 
\label{fig:r_vs_t}}
\end{center}
\end{figure}

 Figure \ref{fig:r_vs_t} shows the location of the outer shock (measured by its peak density) as a function of time for various models 
 (KE, LD, TEa) and SN parameters in absence of cooling.
  The solid line at the bottom shows the transition from a single blast wave (outer shock radius, $r_{\rm OS} \propto t^{2/5}$) to 
 a continuously driven bubble ($r_{\rm OS} \propto t^{3/5}$; \citealt{wea77}) for $N_{\rm OB}=100$ run. The runs with more SNe 
 show such a transition very early on. The dot-dashed line shows the outer shock radius for the KE run using a large ejecta radius violating 
 the criterion in Eq. \ref{eq:r_therm}; the outer shock radius is much smaller than expected  because energy is overwritten
 before it energizes the hot bubble (see section \ref{sec:over}). The luminosity driven (LD) and kinetic explosion (KE) models agree only if the ejecta 
 radius satisfies Eq. \ref{eq:r_therm} for KE models (we have verified that this constraint also applies to the thermal explosion overwrite [TEo] models).
 TEa (thermal explosion addition) runs and LD runs fall on top of each other for both choices of $r_{\rm ej}$ (2, 100 pc).  The outer shock radii  
 for the runs with $r_{\rm ej}=100$ pc increase only after a thermalization time (Eq. \ref{eq:t_therm}; although in this case $\rho$ is not the ISM density 
 but the much lower density of the bubble within which energy is added).

\subsection{CC85 wind within the bubble}
\label{sec:CC85_sims}
\begin{figure}
\begin{center}
\psfrag{B}[cc][][1.][0]{$r({\rm pc})/N_{\rm OB,100}^{1/5}$}
\includegraphics[scale=0.4]{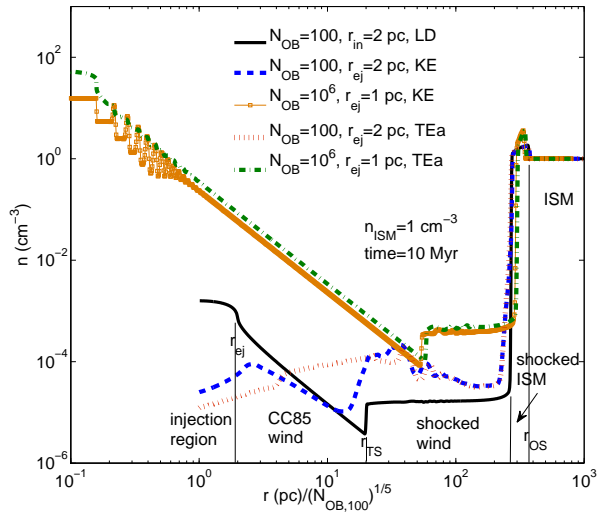}
\caption{Density profile as a function of normalized radius for luminosity driven (LD),  kinetic explosion (KE), and thermal explosion 
addition (TEa) models. The standard CC85 wind within the bubble appears for the LD model, and for KE and TEa models 
with $N_{\rm OB}=10^6$, but not for KE/TEa models with $N_{\rm OB}=100$;  the smooth CC85 wind is identified by the density profile 
varying $ \propto r^{-2}$ between the ejecta radius and the termination shock (various regions have been marked for the LD run). 
The CC85 wind density using $N_{\rm OB}=10^6$ is slightly smaller for the KE model compared to the TEa model because density is overwritten (and 
hence mass is lost) in KE models.  
\label{fig:rho_vs_r_CC85}}
\end{center}
\end{figure}

In this section we show that a simple steady wind, as predicted by CC85, exists within the bubble only if the number of 
SNe is sufficiently large (see section \ref{sec:CC85}). Figure \ref{fig:rho_vs_r_CC85}  shows the density profile as 
a function of the scaled radius for various models.
The solid line shows density for a luminosity driven (LD) model with $N_{\rm OB}=100$ and $r_{\rm ej}=2$ pc; various regions for the smooth CC85 wind within the bubble are marked. The 
superwind has a structure identical to the CC85 wind; the sonic point is just beyond the energy injection radius (2 pc). 
The wind shocks at the termination shock ($r_{\rm TS}$) where the wind ram pressure balances the bubble pressure. 
The ratio of the termination shock and the outer shock ($r_{\rm OS}$) is $\approx 0.07$, in good agreement with Eq. \ref{eq:r_TS}. For 
comparison, Figure \ref{fig:rho_vs_r_CC85} also shows the density profiles for the 
kinetic explosion (KE) and thermal explosion addition (TEa) models with the same parameters. 
While the outer shock radius agree for these runs, the density profiles within the bubble 
are quite different. The most blatant difference, for runs with $N_{\rm OB}=100$, is the absence of a CC85 wind in 
KE and TEa models. In accordance with the discussion in section \ref{sec:CC85}, SN shocks do not thermalize within the
termination shock for a small number of SNe (see Eqs. \ref{eq:rth_CC} \& \ref{eq:CC_crit}); therefore a smooth CC85 
wind is not expected in any model with small $N_{\rm OB}$ except LD.

Only for a large enough $N_{\rm OB}$ and late enough times does a CC85 wind start to appear within the hot bubble. 
Figure  \ref{fig:rho_vs_r_CC85} includes the density profiles for kinetic explosion (KE) and TEa models using 
$N_{\rm OB}=10^6$ (the inner [outer] radius of the computational domain for these runs is 0.5 pc [5 kpc]; $r_{\rm ej}=1$ pc is chosen to satisfy the constraint in Eq. \ref{eq:r_therm}). 
Clearly, in these cases we see the appearance of the CC85 wind solution within the termination shock because the 
injected energy is thermalized. For the KE run with $N_{\rm OB}=10^6$ one can still see 
the internal shocks due to isolated SNe interacting with the superwind.
The density profile for the KE model using $N_{\rm OB}=10^5$ is shown by the dotted line in Figure \ref{fig:rho_vs_r}. 
In agreement with Eq. \ref{eq:r_TS}, the ratio $r_{\rm TS}/r_{\rm OS}$ increases with an increasing $N_{\rm OB}$. 
For $N_{\rm OB}=10^5$ thermalization is less complete as compared to $N_{\rm OB}=10^6$, but happens within
the termination shock. In comparison, a clear termination shock is absent for $N_{\rm OB}=100$ because the thermalization
radius is larger than the termination shock radius (see Eq. \ref{eq:CC_crit}). 
 
 \subsection{Effects of radiative cooling}
  
 In this section we study the effects of radiative cooling on SNe and SBs. We focus on a few aspects:  the fizzling out of 
 thermal feedback in some models in which energy is not injected over a sufficiently small scale; 
 comparison of cooling losses  and mechanical energy retained by radiative SNRs and SBs; the influence of 
 magnetic fields and thermal conduction.
 
 \subsubsection{Unphysical cooling losses with thermal energy addition}
\label{sec:art_loss}
As we mentioned in section \ref{sec:cool_cond}, 
some models (TEa, LD) in which we {\em add} SN thermal energy in a dense ISM, over a large radius, 
can suffer unphysical catastrophic radiative cooling. 
In such cases a hot bubble is not even created and SN feedback has no effect, whatsoever. 
Early SN feedback simulations suffered from this problem because of low resolution. 

\begin{figure}
\begin{center}
\includegraphics[scale=0.5]{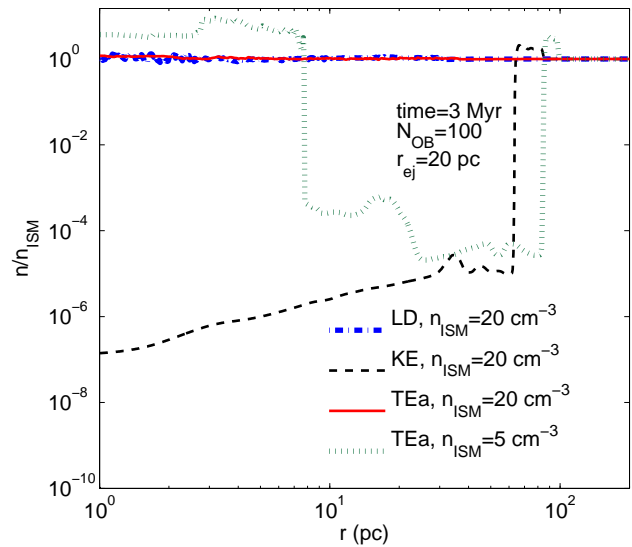}
\caption{Density as a function of radius for different runs at 3 Myr to show that energy addition totally fizzles out for a high ISM density.
While TEa and LD models do not show the formation of a hot, dilute bubble for ISM density of 20 cm$^{-3}$, KE model indeed shows 
a bubble and a forward shock. Also shown is the density profile for TEa model with a lower density (5 cm$^{-3}$) ISM;  at later times 
it shows a bubble which pushes the shell outwards.  The outer shock radius is larger for a lower density ISM because 
$r_{\rm OS} \propto \rho^{-1/5}$.
\label{fig:fizzling}}
\end{center}
\end{figure}

Figure \ref{fig:fizzling} shows the density profiles at 3 Myr for three of our energy injection models (KE, LD, TEa) with $N_{\rm OB}=100$ and 
the ISM density of 20 cm$^{-3}$.
The ejecta radius is chosen to be large such that it  violates conditions in Eqs. \ref{eq:lum_fizzle} \& \ref{eq:eadd_fizzle}. 
The figure shows a comparison of the luminosity driven (LD) and thermal explosion addition (TEa) models that fizzle out, and kinetic explosion (KE) model which shows a hot, dilute 
bubble.  Thus, our results are in agreement with the
analytical considerations  of section \ref{sec:cool_cond}. The outer shock location for the KE model roughly agrees with the self similar scaling of \citet{wea77} 
if the luminosity is reduced by a factor $\approx 0.35$; this is comparable to the fraction of mechanical energy retained by SBs after the outer shock becomes radiative (see section \ref{sec:energy_budget} and the right panel of Fig. \ref{fig:rad_loss_time}).

For most runs in Figure \ref{fig:fizzling} we have chosen a rather high density ($n=20$ cm$^{-3}$) compared to 
the critical values in Eqs. \ref{eq:lum_fizzle} \& \ref{eq:eadd_fizzle}. For lower densities (e.g., 5 cm$^{-3}$ for TEa model in Fig. \ref{fig:fizzling}) 
we find that the energy does not 
couple at early times. Energy injection excites large amplitude sound waves and associated density perturbations, such that at late times the lowest density 
regions no longer violate Eqs. \ref{eq:lum_fizzle} \& \ref{eq:eadd_fizzle}. After this, a hot bubble starts to grow because of energy injection in a 
dilute medium (see dotted line in Fig. \ref{fig:fizzling}), and eventually the outer shock radius starts to agree with analytic estimates.

 \subsubsection{SB evolution with cooling}

This and later sections, which study the influence of radiative cooling on SBs and SNRs, use the realistic kinetic explosion (KE) model
for supernova energy injection with ejecta radius $r_{\rm ej}=2$ pc. However, we have verified that other models discussed in section 
\ref{sec:fb_pres} give similar results, as long as the conditions in section \ref{sec:anal} are satisfied.

Spherical adiabatic blast waves, both SNRs and SBs, have shells with finite thickness. An estimate 
for the shell thickness is obtained by assuming that all the swept-up ISM mass lies in a shell and that the post shock density 
is 4 times the ISM density for a strong shock; this gives $\Delta r/r_{\rm OS} \approx 1/12$. Of course, the shock transition 
layer is unresolved in simulations, and in reality is of order the mean free path.
The structure of an adiabatic blast waves is fairly simple. The density jump at the shock is 4 for a strong shock, and as the shock 
becomes weaker the density jump decreases and the shell becomes broader. Eventually, the outer shock is 
so weak that it no longer compresses gas irreversibly, but instead becomes a sound wave with compressions and 
rarefactions (see Fig. 2 in \citealt{tan05}). 

Since the evolution of isolated SNRs with cooling has been thoroughly studied in past (e.g., \citealt{tho98}; hereafter T98), we only highlight 
the differences between isolated SNRs and  SBs. The fundamental difference between the two is that SNRs suffer
catastrophic losses just after they become radiative because, unlike in SBs, there is no energy injection after this stage. 
In SBs, the cool (yet dilute), fast SN ejecta periodically thermalizes within the bubble 
and powers it long after the forward shock 
becomes radiative. This keeps the radiative forward shock moving (like a pressure-driven snowplow), as long as SNe go off within 
the hot bubble.

\begin{figure*}
\begin{center}
\includegraphics[scale=0.55]{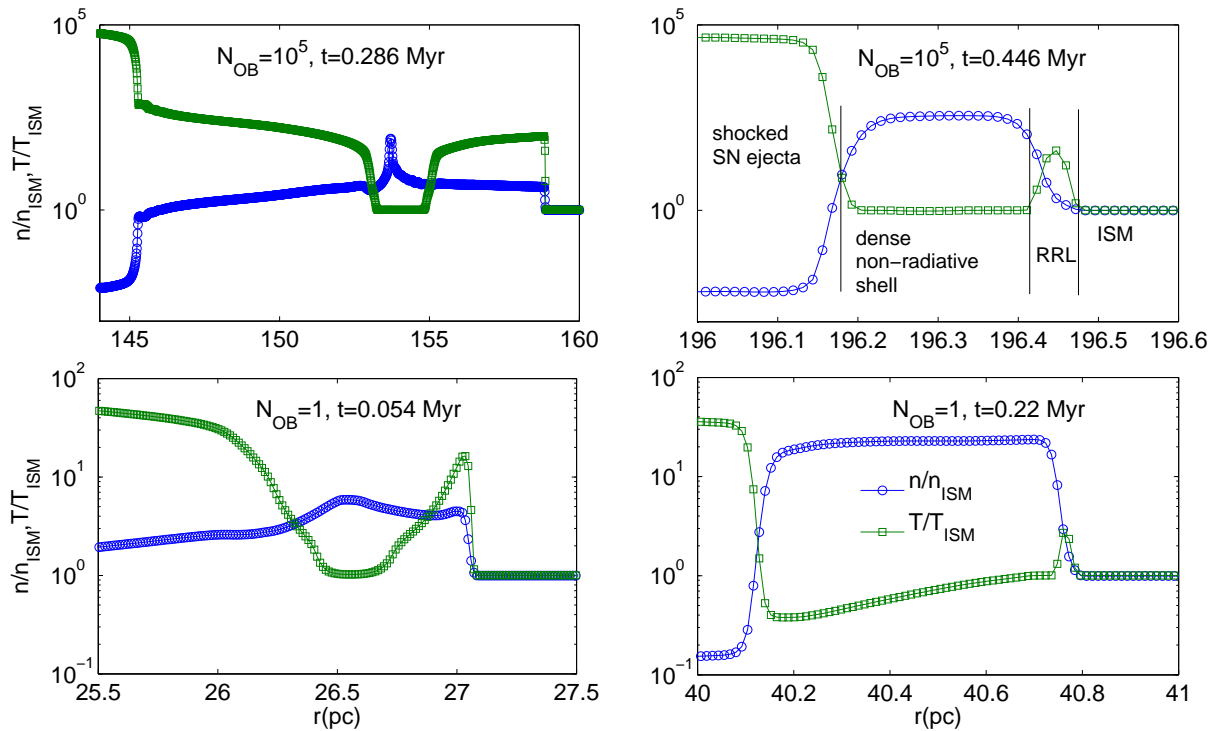}
\caption{The normalized (with respect to the ISM) density and temperature profiles zoomed-in on the outer shock as a function of radius for the high 
resolution (16384 grid points uniformly spaced from 1 to 200 pc) runs. Top panel: $N_{\rm OB}=10^5$ run; bottom panel: a 
single SNR ($N_{\rm OB}=1$) run. Left panels correspond to a time when the outer shocks just become radiative 
and the right panels are for later times. Markers represent the grid centers. For a single SNR the temperature 
in the dense shell is lower than the temperature floor (ISM temperature) because of weakening of the shock and the resultant 
adiabatic losses. Different regions (unshocked ISM, radiative relaxation layer, dense non-radiative shell, 
and shocked SN ejecta) are marked in the top-right panel.
\label{fig:d_T}}
\end{center}
\end{figure*}

The structure of a radiative shell is quite complex. The shell become radiative when the cooling time of the post-shock gas is 
shorter than its expansion time (which is of order the age of the blast wave). 
Moving inward from the upstream ISM, the outer shock transition happens over a mean free path, which is followed 
by a thin radiative relaxation layer of order the cooling length (see, for example,  pp. 226-229 of \citealt{shu92} and
the top-right panel of Fig. \ref{fig:d_T}). The
 radiative relaxation layer is followed by a  dense shell, which is separated by a contact discontinuity  from the dilute
 hot bubble.  In steady state, radiative cooling is 
concentrated at two unresolved boundary 
layers, the outer radiative relaxation layer and the inner contact discontinuity. Here the density is high and 
the temperature is conducive to radiative cooling.

 Figure \ref{fig:d_T} shows the zoomed-in density and temperature structure of the radiative shell for a SB (with $N_{\rm OB}=10^5$; upper panels) 
 and a SNR 
 ($N_{\rm OB}=1$; lower panels) using high resolution (with 16384 grid-points) runs. It clearly shows an outer radiative shock and an inner contact discontinuity. 
 Within the contact discontinuity of the SB ($N_{\rm OB}=10^5$) is the shocked SN ejecta; 
Figure \ref{fig:rho_vs_r_CC85} shows the full structure of a superwind within the SB. Just when the outer shock becomes radiative the coolest/densest part is compressed by high pressure regions sandwiching  it (left panels of Fig. \ref{fig:d_T}). 
After a sound crossing time the post-shock region is roughly isobaric and in the pressure-driven snowplow phase (right panels 
of Fig. \ref{fig:d_T}). 

Unlike SBs, for isolated SNRs there is no energy injection at later times; the pressure in the bubble falls precipitously after
the outer shock becomes radiative at $\approx$0.05 Myr. By $\sim 0.5$ Myr the bubble pressure 
becomes comparable to the ISM pressure, the shell density falls and it becomes momentum conserving with a velocity 
comparable to the sound speed in the ISM. At even later times ($\sim$ few Myr) the hot bubble just oscillates as a weak acoustic wave.

\begin{figure}
\begin{center}
\includegraphics[scale=0.43,angle=0]{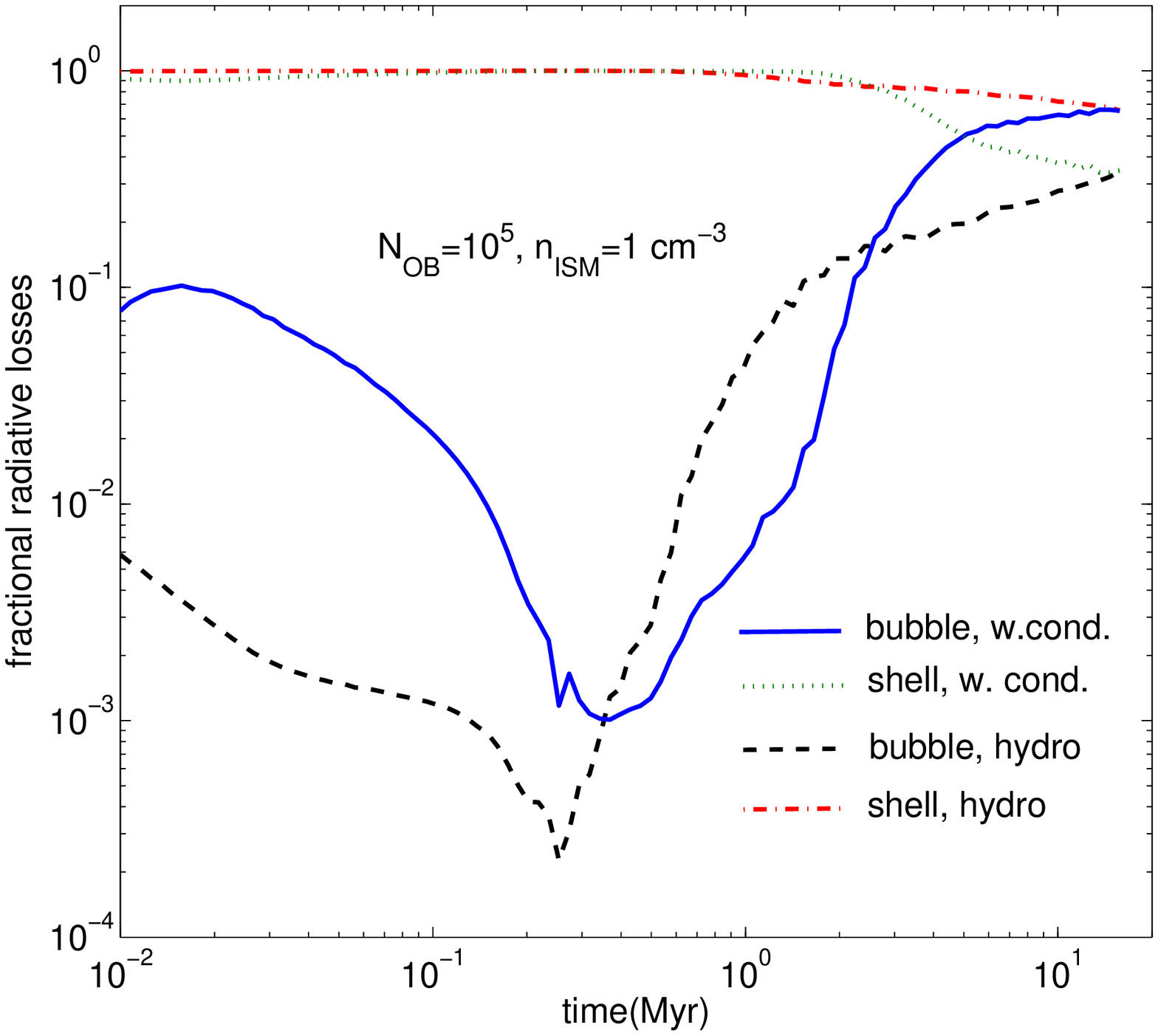}
\caption{Fractional radiative losses in shell ([shell cooling rate]/[total cooling rate]) and bubble ([bubble cooling rate]/[total cooling rate]) 
for KE models  ($N_{\rm OB}=10^5$) with and without conduction ( the run with thermal conduction is discussed in section \ref{sec:B_cond}). Most radiative 
energy losses happen at the radiative relaxation layer ahead of the dense shell. At late times, as the outer shock weakens, radiative losses 
in the bubble become more dominant. Bubble is comparatively more radiative (in fact, bubble losses exceed shell losses after 5 Myr) 
with conduction because of mass loading of the bubble by evaporation from the dense shell. Results from the high resolution run and the
luminosity driven (LD) model are similar. The minimum in fractional radiative losses corresponds to the time when the outer shock 
becomes radiative.
\label{fig:rad_ratio}}
\end{center}
\end{figure}

Figure \ref{fig:rad_ratio} shows the distribution of radiative losses in the shell and in the bubble (shell is defined as the outermost region where density is above 1.01 times the ISM density; bubble comprises of all the grid points with radius smaller than the inner shell radius)  for a superbubble with $N_{\rm OB}=10^5$; results from runs with and without thermal conduction are shown. Here we only discuss the run without conduction; the run with conduction is highlighted in section \ref{sec:B_cond}. 
Unlike Figure \ref{fig:d_T}, here we use our standard resolution runs (1024 grid points) because we are running for a much longer time. Results 
from the higher resolution runs match our standard runs, highlighting the fact that the volume integrated cooling is the same even if the radiative relaxation layer and the contact discontinuity are unresolved. The time and volume integrated losses ($\int \int n^2 \Lambda 4 \pi r^2 dr dt$) in
 bubble and shell are sampled appropriately and differentiated in time to obtain their respective cooling rates.
 As already discussed, cooling is concentrated at the radiative relaxation layer, which is included in the shell, and at the contact discontinuity, 
 part of which is included in the bubble. Consistent with our previous discussion, most  radiative losses are concentrated in the shell. Fractional radiative losses in the bubble (concentrated at the contact 
 discontinuity) are $\sim 10^{-4}-0.3$, which increase with time as the outer shock becomes weaker. 
 
\subsubsection{Scales in radiative shocks}
The structure of a radiative shock is well known (see, e.g., pp. 226-229 in \citealt{shu92}). Applying mass and momentum 
conservation across the radiative relaxation layer in the shock frame, and assuming the same temperature upstream/downstream of it 
(see the top-right panel of Fig. \ref{fig:d_T} for different regions of the outer shock) gives $\rho_3/\rho_1 = (u_1/u_3) = (u_1/a_T)^2$ 
(Eq. 16.36 in \citealt{shu92}), where $\rho_3~(\rho_1)$ is the density downstream (upstream) of the radiative relaxation 
layer, $u_1$ ($u_3$) is the upstream (downstream) velocity in the shock rest frame, and $a_T$ is the isothermal sound speed
of upstream ISM (at $T=10^4$ K below which radiative cooling vanishes). Thus, we expect larger density jump across stronger shocks 
($u_1 \gg a_T$). 
This is evident from the shell density for the two cases ($N_{\rm OB}=10^5,~1$) in Figure \ref{fig:d_T}. 

The thickness of the cold, dense shell can be estimated 
by equating the swept-up ISM mass with the mass in the constant density shell; $\Delta r/r_{\rm OS} \approx (a_T/u_1)^2/3$. 
This thickness is quite small, with $\Delta r/r_{\rm OS} \approx 0.003$ for a 100 km s$^{-1}$ shock. This estimate agrees with our results 
in Figure \ref{fig:d_T}, and as predicted, the shell is thicker for a smaller $N_{\rm OB}$ and becomes thicker with time as the shock becomes 
weaker.  

The thickness of the radiative relaxation layer can also be estimated. The size of the radiative relaxation layer is $L_{\rm cool}$  ($L_{\rm cool}$ 
is the distance behind the outer shock after which the advection time becomes longer than the cooling time), such that
\be
\label{eq:RL}
\int_0^{L_{\rm cool}} \frac{dx}{u} = \int_0^{t_{\rm cool}} dt = t_{\rm cool},
\ee
where $u(x)$ is the velocity in the relaxation layer in the shock rest frame. While this equation can only be solved after numerically 
solving for the structure of 
 the relaxation layer, we can make an order of magnitude estimate. The integral on the LHS of Eq. \ref{eq:RL} can be estimated 
as $L_{\rm cool}/\la u \ra$, where $\la u \ra = a_T/2$ is the geometric mean of the velocity at the front of the relaxation layer 
($u_1/4$ for a strong shock) and just downstream of it ($u_3 = a_T^2/u_1$). Similarly, the cooling time $t_{\rm cool}$ in Eq. \ref{eq:RL} 
can be estimated by using a geometric mean of densities across the relaxation layer; i.e., 
$t_{\rm cool} \approx 1.5 k T/(\la n \ra \Lambda) $, where $\la n \ra = 2 (u_1/a_T) n_1$ and we can use the peak of the cooling 
function for $T$ and $\Lambda$. Putting this all together gives,
\be
L_{\rm cool} \sim a_T \left ( \frac{a_T}{u_1} \right ) \frac{k T_0}{n_1 \Lambda_0},
\ee
which is $\sim 10^{-4}$ pc for fiducial numbers, far from being resolved even in our highest resolution runs. While the transition 
layers (contact discontinuity and radiative relaxation layer) where all our cooling is concentrated are unresolved, we find that 
the volume integrated quantities such as radiative losses, kinetic/thermal energy in shell/bubble are converged even at our 
modest (1024 grid points; results are similar even for 256 grid points) resolution. 

\subsubsection{Energetics of radiative SBs \& isolated SNRs}
\label{sec:energy_budget}

\begin{figure}
\begin{center}
\includegraphics[scale=0.5,angle=0]{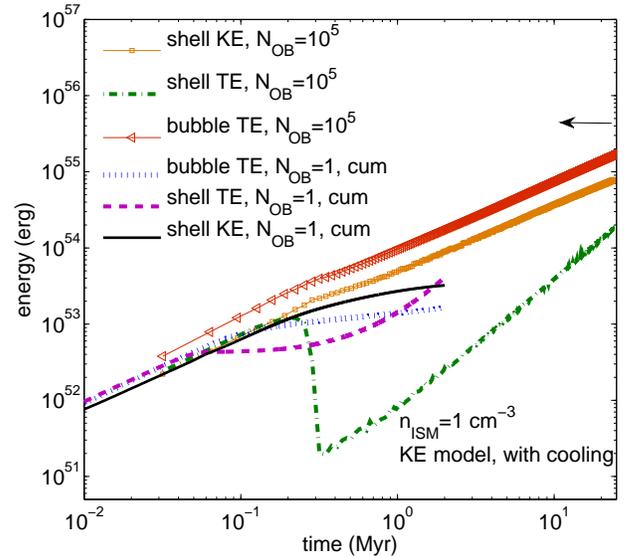}
\caption{Comparison of kinetic and thermal energies in the shell and thermal energy in the bubble as a function of time for SBs and an equal number of isolated 
SNe. Results from an isolated SN run ($N_{\rm OB}=1$) have been combined cumulatively (see Eq. \ref{eq:cum}), assuming that SNe go off 
independently in  the ISM. Pre-radiative phase energetics are similar but isolated SNRs are extremely deficient 
in mechanical energy (after 1 Myr) as compared to a SB with the same energy input. The arrow on top right shows the bubble thermal energy
at the end for an adiabatic SB run. Isolated SN results are only shown till 2 Myr because SNRs become weak sound waves by then. 
\label{fig:energy}}
\end{center}
\end{figure}

In this section we focus on the energetics of SB shell/bubble and compare it with the results from isolated SNRs. We define the 
shell to be the outermost region where the density is larger than 1.01 times the ISM density. All gas at radii smaller than the shell inner radius 
is included in the bubble (this definition is convenient but not very precise as it includes small contribution from the unshocked SN ejecta). Figure
  \ref{fig:energy} shows a comparison of kinetic and thermal energies in bubble and shell as a function of time for a SB driven by $10^5$ SNe. 
  The bubble kinetic energy is not included because it is much smaller. Also included is a comparison of the same quantities for the 
  same frequency of SNe that go off independently. The results for multiple isolated SNe are obtained by combining the single SN run 
  at different times. We simply use the data at an interval of 
  $t_{\rm SN}$ (time between individual SNe) and add them to obtain total kinetic/thermal energy in the shell and bubble at a given time.
  For instance, the thermal energy in bubbles of all independent SNe at time $10t_{\rm SN}$ is obtained by summing up the bubble 
  thermal energy from a single SN ($N_{\rm OB}=1$) run at $t=0$, $t_{\rm SN}$, $2t_{\rm SN}$, ..., till $10t_{\rm SN}$. This is equivalent 
  to a cumulative sum over time for a single SN run,
  \be
  \label{eq:cum}
E_{\rm cum}(t) = \sum_{i=0}^{i<N} E(i t_{\rm SN}) =  \frac{1}{t_{\rm SN}}\int_0^t E(t^\prime) dt^\prime,
  \ee
  where $E$ stands for, say, bubble thermal energy and $N$ is the number of SNe till time $t$.

 \citet{wea77} have given analytic predictions for energy in different components of SBs: the total energy 
 of the shell is $(6/11)L_{\rm ej}t$ ($40\%$ of this is kinetic energy and $60\%$ is thermal) and the thermal energy of the bubble is 
 $(5/11)L_{\rm ej}t$ (kinetic energy of the bubble is negligible). These analytic predictions agree well with our numerical results in the 
 early adiabatic (non-radiative) SB phase in Figure \ref{fig:energy}. 

Figure \ref{fig:energy} shows that the SB shell loses most of its thermal energy catastrophically at $\approx 0.25$ Myr; the 
trough in shell thermal energy can be estimated by assuming that all the swept-up mass till then cools to the stable temperature ($10^4$ K). The 
thermal energy of the cold shell increases after that as it sweeps up mass from the ISM; this is not a real increase in the thermal energy because the newly added material, which was previously part of the ISM, simply becomes a part of the dense shell at the same temperature. 
The bubble thermal energy and the shell 
kinetic energy show only a slight decrease in slope after the radiative phase because they are energized by the non-radiative 
termination (internal) shock(s) driven by SN ejecta. However, there are some losses because of cooling at the contact discontinuity (see Figs. \ref{fig:d_T} 
\& \ref{fig:rad_ratio}). 
 
The shell kinetic energy and the bubble thermal energy in radiative SB simulations at 20 Myr are roughly half of the values obtained in 
adiabatic simulations (which agree with analytic predictions). Thus, the mechanical energy retained in the SB is
$\approx 0.34 L_{\rm ej} t$. This should be contrasted with the energy evolution in isolated SNRs. The isolated SNR
becomes radiative much earlier ($\approx$ 0.05 Myr; when the shell thermal energy shown by dashed line flattens suddenly in Fig. \ref{fig:energy}) 
because of a weaker shock compared to a SB. Note that the energies for isolated SNe in Figure \ref{fig:energy} are {\em cumulative sums} 
over time of a single SN run (see Eq. \ref{eq:cum}). 
The bubble thermal energy and shell kinetic energy also drop for an individual SNR after it becomes radiative (albeit not catastrophically, unlike the
shell thermal energy; see Fig. 3 in T98; this is the pressure-driven snowplow stage) because of cooling at the contact discontinuity and adiabatic 
losses, and because there is no new energy source (unlike the termination/internal shocks in a SB). The total mechanical energy in the bubble and shell of a single SNR
at the beginning of the momentum conserving phase (1 Myr; when bubble pressure is comparable to the ISM pressure) is $10^{50}$ erg, which is only 10\% of the input energy 
(see Fig. 3 in T98). This agrees with the energy fraction available as mechanical energy of the SNR, as quoted by T98. After a few Myr the SNR should be considered a non-energetic part of 
the ISM, as the thermal energy of the swept up ISM becomes larger than the SNR's  mechanical energy, and the bubble becomes a weak acoustic wave.

In order to compare isolated SNRs and superbubbles over the cluster lifetime we must extrapolate our cumulative SN energies to 30 Myr.
This is also the relevant timescale for preventing large-scale galactic inflows
from efficiently forming stars (the free-fall timescale at $\sim 10$ kpc for galactic halos is few 10s of Myr). For isolated SNRs the shell kinetic energy + bubble thermal 
energy is  $\sim$7\% of the input energy by 2 Myr, and only 0.7\% when extrapolated to 30 Myr. We should not extrapolate the shell thermal energy 
 because its rise at late times in Figure \ref{fig:energy} is due to the sweeping up of the ISM into the shell, without an increase in the temperature. To conclude, isolated SN
feedback is much weaker (by a factor $\sim 50$) as compared to the feedback due to SBs over the cluster lifetime.

 \begin{figure*}
\begin{center}
\includegraphics[scale=0.5]{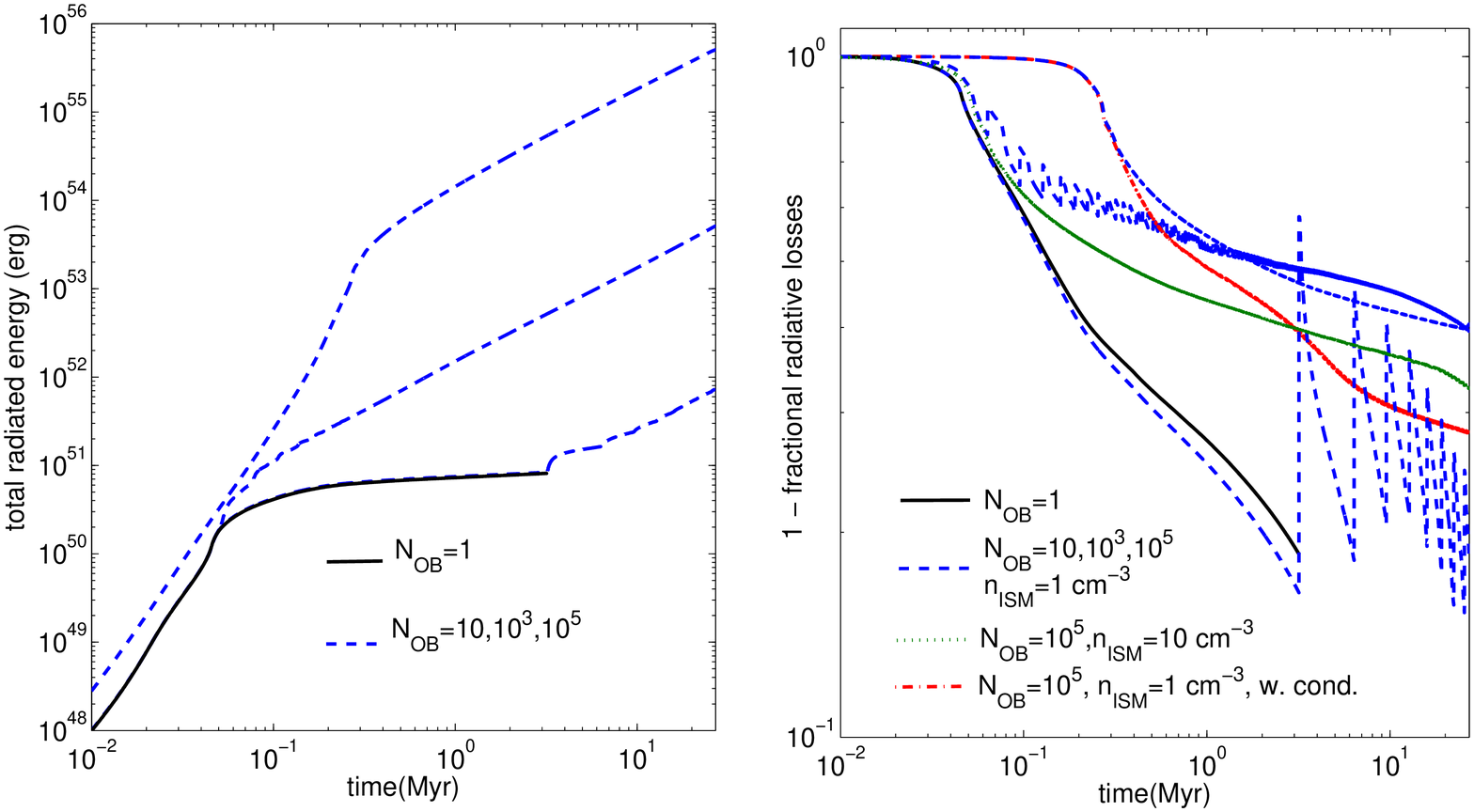}
\caption{Radiative losses as a function of time for SBs and isolated SNe. Left panel shows the total radiated energy as a function of time for an
isolated SN run (solid line) and for SB runs (dashed lines) with $N_{\rm OB}=10,~1000,~10^5$;  larger 
$N_{\rm OB}$ leads to larger radiative losses because of a higher density and temperature in the radiative relaxation layer (see Fig. 
\ref{fig:d_T}). The right panel shows fractional cooling losses (1 - [energy radiated]/[input energy]) as a function of time; the total energy  input at some time equals the number of SNe 
put in by that time multiplied by $10^{51}$ erg (the spikes for $N_{\rm OB}=10,~10^3$ in the right panel reflect the discreteness of SN energy input within SBs). All SB runs, including those with conduction 
and with higher density, show that only a factor of $0.6-0.8$ is radiated by 20 Myr (and a factor of $0.2-0.4$ is retained as mechanical energy). 
In contrast, the isolated SN run (solid line) loses 80\% of its energy by 3 Myr, after which it is no longer over-pressured with respect to the ISM.
\label{fig:rad_loss_time}}
\end{center}
\end{figure*}

The left panel of Figure \ref{fig:rad_loss_time} shows the total radiated energy over the whole computational domain as a function of time 
for an isolated SNR ($N_{\rm OB}=1$; solid line) and for 
SBs (dashed lines). The results are qualitatively different for SBs (even for $N_{\rm OB}=10$) and isolated SNe. While an isolated SN radiates
almost all of the input energy ($10^{51}$ erg) over a Myr timescale, SBs radiate a smaller fraction ($0.6-0.8$) of their energy even till 
late times. The runs with smaller $N_{\rm OB}$ and larger density become radiative at an early time as the shock is weaker, but SB runs are 
qualitatively similar. Unlike in
isolated SNe, a significant fraction of the input energy in SBs is retained in the bubble thermal energy and the shell kinetic energy ($0.2-0.4$;
see Fig. \ref{fig:energy}). The key reason for the difference between isolated SNe and SBs is that in SBs the non-radiative termination/internal  
shocks keep the bubble overpressured but isolated SNe, which do not have further energy input after the initial explosion, simply fizzle out soon after
they become radiative. 

The right panel of Figure \ref{fig:rad_loss_time} shows the fraction of energy retained (1-the fraction radiated) as a function of time 
for several runs. All SB runs, including a higher density run ($n_{\rm ISM}=10$ cm$^{-3}$) and the run with conduction (see section \ref{sec:B_cond}), show the asymptotic 
fraction of energy retained to be $\gtrsim 0.25$. In contrast, an isolated SN loses 90\% of the input energy by few Myr (and almost all of it by 10 Myr; see also Fig. \ref{fig:energy}).
Radiative losses for an isolated SNR at late times ($\gtrsim$ few Myr) are more than the energy input ($10^{51}$ erg); these come at the expense of the thermal
energy of the swept-up ISM.

We can compare our results of coincident SNe with the case of multiple SNe distributed over space in a random manner. The two cases presented in Figure 
\ref{fig:energy} represent two extreme limits: spatially coincident SNe in a SB and totally independent SNe. For spatially distributed SNe we expect 
results somewhere in between these two extremes. \citet{vas14} compare the total 
explosion energy that remains as the thermal energy of hot gas in the case of spatially distributed SN explosions. They study the effects of coherent 
explosions, as defined in \citet{roy13}, which implies that  SNe overlap before they become radiative.
If the shell radius of a SNR when it becomes radiative is $R_a$ and the corresponding time scale is $t_a$, then for a 
SN rate density of $\nu_{\rm SN}$, the coherency condition is that $(4 \pi/3) \, R_a^3 \, t_a \, \nu_{\rm SN} > 1$. \citet{vas14} compare the cases in 
which explosions occur coherently with those in which they do not. They find that a fraction $\sim 0.3$ of the explosion energy is retained in the 
gas with temperature $T \ge 3 \times 10^6$ K if the explosions occur coherently, and the fraction is $0.02-0.2$ if the explosions 
are incoherent. Our results here for SBs correspond to the coherent case, since $t_{\rm SN}$ is always shorter than the cooling time of the gas in the 
bubble. Therefore, our result of a fraction $\sim 0.35$ being retained as SB's mechanical energy is consistent with \citet{vas14}. 

\subsubsection{Effects of magnetic fields and thermal conduction}
\label{sec:B_cond}
\begin{figure*}
\begin{center}
\includegraphics[scale=0.45]{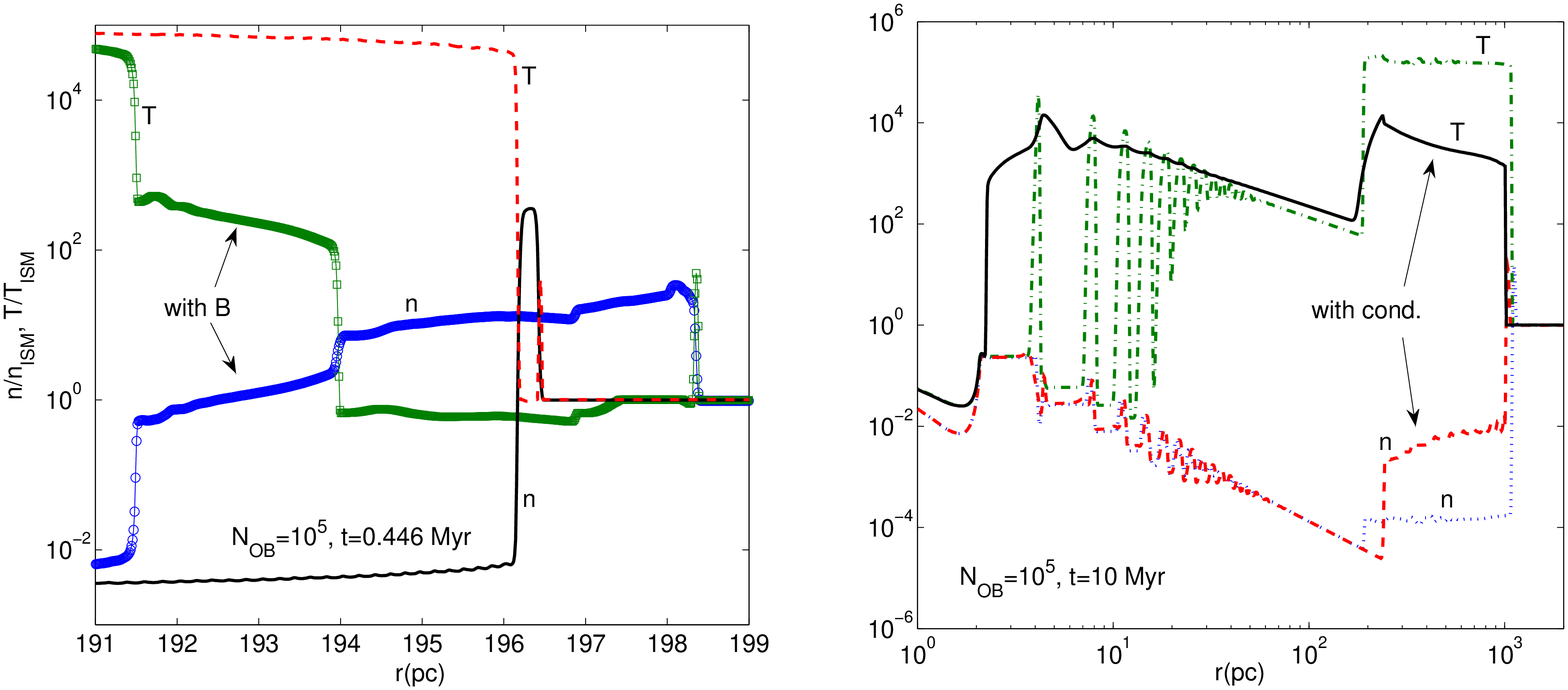}
\caption{The normalized density and temperature profiles to show the effects of magnetic fields and thermal conduction on SB evolution with cooling. The left panel shows the profiles zoomed in
on the outer shock for MHD (initial $\beta=1$) and hydro runs with 16384 grid points. Magnetic field is enhanced in the shell and the shell is thicker. The right panel shows the
profiles for radiative hydro runs with and without thermal conduction (1024 grid points); unlike in the left panel, we show the whole computational domain and the dense shell is barely visible. Thermal conduction evaporates mass from the dense shell and spreads it into the bubble, thereby 
making it denser and less hot compared to the hydro run. The temperature structure in the internal shocks (within the superwind) is also smoothened out by thermal 
conduction.
\label{fig:dT_B_cond}}
\end{center}
\end{figure*}

Since the ISM is magnetized, we try to assess the qualitative effects of magnetic fields using an idealized high resolution (16384 grid points) MHD simulation. We assume 
an azimuthal ($\phi$) component of the 
magnetic field so that only magnetic pressure forces (and no tension) are present. We choose a plasma $\beta$ (ratio of gas pressure and magnetic pressure) 
of unity in the ISM, and our SN ejecta is also magnetized with the same value of $\beta$. Since the ejecta is dominated by kinetic energy and the bubble is 
 expanding, we do not expect magnetic fields to affect the bubble and the ejecta structure. However, the radiative shell is compressed because 
of cooling, and due to flux-freezing magnetic pressure is expected to build up in the dense shell. This is indeed what we find in our simulations with magnetic 
fields. The left panel of Figure \ref{fig:dT_B_cond} shows the zoomed-in density and temperature structure of the radiative outer shock with and without magnetic fields.
The key difference between the hydro and MHD runs is that the dense shell in MHD has a lower density and is much broader. This is because magnetic pressure
prevents the collapse of the dense shell.\footnote{The photon mean-free-path for a dense shell can become smaller than the shell thickness. When this happens, the assumption of optically thin cooling breaks down, and the shell can become thicker because of radiation pressure.} The dense shell ($194~{\rm pc}<r<198~{\rm pc}$ in the left panel of Fig. \ref{fig:dT_B_cond}) is magnetically dominated with 
plasma $\beta \sim 0.01$. The MHD run has
two contact discontinuities; one at the boundary of the hot bubble ($r \approx 191.5$ pc), and another at $r \approx 194$ pc left (right) of which 
the plasma is dominated by thermal (magnetic) pressure. 

Another important physical effect, especially in the hot bubble is thermal conduction. We carry out a 1024 resolution hydro run with thermal conduction to study its 
qualitative influence. However, it is difficult to determine the ISM conductivity in a magnetized  (presumably 
turbulent) plasma. Therefore, we use the Spitzer value with a suppression factor of 0.2 (see Eq. 11 in \citealt{sha10}). Moreover, since the bubble can become very hot such that the diffusion approximation 
breaks down, we limit the conductivity to an estimate of the free streaming diffusivity (chosen to be 2.6 $v_t r$ where $v_t$ is the local isothermal sound speed and $r$ is the 
radius). Thermal conduction is operator split, and implemented fully implicitly through a tridiagonal solver using the code's hydro time step.

Conduction is expected to evaporate matter from the dense shell and deposit it into a conductive layer in the bubble; in steady state the rate of conductive transport 
of energy from bubble to the shell is balanced by the rate of heat advection from shell to the bubble \citep[][]{wea77}. The outer and termination shock locations are not 
affected much by conduction. However, the density and temperature structure in the hot bubble is affected significantly. Without conduction the bubble is very hot 
($\sim 10^9$ K), but with conduction the temperature drops into the X-ray range ($10^7-10^8$ K) and density is higher. This can enhance the X-ray emissivity 
of SBs; 
a rough estimate of hard X-ray luminosity ($\int 4 \pi  r^2 n^2 \Lambda dr$ over the hot bubble) at 10 Myr from the right panel of Figure \ref{fig:dT_B_cond} is few $10^{38}$ erg s$^{-1}$ (which is $\sim 0.003$ the energy put in by SNe by that time). Since galactic superwinds are copious X-ray emitters, we expect thermal conduction to be a very important ingredient for explaining observations. Figure \ref{fig:rad_ratio} confirms that the fraction of radiative losses from the bubble is much higher with conduction than without conduction because of a higher density. However, the right panel of Figure \ref{fig:rad_loss_time} shows that the total fractional radiative losses with thermal conduction are only slightly higher compared to the non-conductive $N_{\rm OB}=10^5$ SB run.

We emphasize that our treatment of magnetic fields and thermal conduction is extremely simplified. Realistic calculations must be done in three dimensions with tangled magnetic fields and with anisotropic thermal conduction along fields lines. However, we expect the qualitative effects of realistic magnetic fields and thermal conduction to have some semblance with our simplified treatment.

\section{Conclusions \& astrophysical implications}
\label{sec:conc}
We have obtained several important results in this paper, on both numerical implementation of SN/SB feedback and on differences between isolated 
SNe and SBs. SBs are a result of spatially and temporally correlated SNe. Since most massive stars are expected to be born in star clusters 
few 10s of pc in extent (e.g., \citealt{lar99}), pre-SN stellar winds and first SNe are expected to carve out a low density bubble, which by a fraction of Myr 
encloses the whole star cluster (see Fig. \ref{fig:d_T}). Therefore, subsequent SNe happen in the low density bubble and we are in the SB regime
of coherent SNe (see \citealt{roy13,vas14}). 

Magnificent galactic outflows, such as M82, are powered by multiple super-star-clusters and the problem of understanding coalescing SBs is important.
Star clusters more massive than $10^5 \msun$ (and hence with $>1000$ SNe) are rare (e.g., see \citealt{pro10}); therefore, the superbubbles in M82 and in our $N_{\rm OB}=10^5$ SB model should be considered as giant bubbles driven by 100s of overlapping superbubbles  due to individual star clusters. Indeed, 100s of star clusters have been observed in the central few 100 pc of M82 (\citealt{oco95}). We note that  vertical stratification is important for the acceleration and assimilation of the metal-rich bubble into the halo. In this paper we consider the idealized smaller-scale problem of the 
behavior of isolated SNRs and multiple coincident SNe within a SB in a uniform ISM. Some of our most important results are: 

\begin{itemize}

\item Our most realistic kinetic explosion (KE) models (and other models in which SN energy is {\em overwritten}), in which the SN energy in kinetic form is overwritten in a small volume, give correct results only if the energy is deposited within a small length scale (see section \ref{sec:over}); otherwise, energy is overwritten without coupling to the ISM.  This is true even without considering any radiative losses.

\item With cooling, if feedback energy is deposited within a length scale $r_{\rm ej}$ larger than the critical values mentioned in section \ref{sec:cool_cond}, such that the input energy is radiated before it is thermalized, a hot bubble is not formed in the widely used luminosity driven and thermal explosion addition (LD, TEa) models. Thus SN fizzles out at early stages due to artificially large cooling losses. 
 
 \item With insufficient resolution and large ISM densities the bubble fizzles out completely in the luminosity driven (LD) and thermal explosion addition (TEa) models (in which energy is {\em added} to the ISM; see section \ref{sec:fb_pres}) and cannot have any effect on the ISM (see Fig. \ref{fig:fizzling}).  As also pointed out previously (e.g., by \citealt{cre11,dal12}), early galactic-scale SN feedback simulations failed mainly because of this. However, for a realistic SN (as mimicked by our KE model) a bubble is formed and subsequent SNe occuring within the non-radiative bubble power the radiative outer shock. Another, probably more serious, problem faced by numerical simulations is that the SN energy is not typically put in coherently over a small volume in space and within a short interval. Feedback due to SNe in young star clusters is expected to be coherent and much more effective than a similar number of isolated SNe (see Figs. \ref{fig:energy}, \ref{fig:rad_loss_time}). For correlated SNe, isolated SN bubbles overlap to form a superbubble which is powered by the non-radiative termination/internal shocks, long after the outer shock becomes radiative. In contrast, an isolated supernova becomes powerless a bit after ($\sim 1$ Myr) the outer shock becomes radiative.
\item A smooth CC85 wind within the superbubble is possible only if the number of SNe ($N_{\rm OB}$) over the cluster lifetime is large (i.e., $N_{\rm OB} \gtrsim 10^4$). Only in these cases, individual SNe going off inside the superbubble are able to thermalize within the termination shock. This result has implications for modeling the X-ray output, for example, in individual bubbles blown by star clusters and in the inner regions of galactic outflows, since the CC85 wind structure is often assumed where it may not be valid.

\item Most of the radiative losses come from the unresolved radiative relaxation layer at the outer shock. The fractional radiative losses from the interior region, concentrated at the contact discontinuity between the shocked ISM and the shocked ejecta, varies between $\sim 0.001-0.3$, with larger losses occurring at later times. While these radiative layers are unresolved even in our highest resolution simulations, the volume integrated radiative losses in them converge even for a modest resolution.

\item As compared to isolated SNe, superbubbles can retain a larger fraction of the initial energy of explosions as thermal/kinetic energy of the gas. Isolated SNe are mixed with the ISM soon after they become radiative; by few Myr they are incapable of affecting the ISM at all. While most energy is radiative away (close to 100\%, and not $\sim 90$\%, as is often assumed) for isolated SNe over 10 Myr, a SB can retain a fraction  $\sim 0.35$ (for $n=1$ cm$^{-3}$) as the bubble thermal energy + the shell kinetic energy. This fraction is only weakly affected by a higher ISM density and by thermal conduction (see the right panel of Fig. \ref{fig:rad_loss_time}). Thus, SBs are expected to significantly affect even a dense ISM. Substantial radiative losses can partly explain the smaller observed bubble sizes compared to what is expected by modeling the stellar populations (see \citealt{oey09} for a summary).

\item The temperature profile of SBs strongly depend on thermal conduction, whose inclusion can decrease (increase) the temperature (density) and thereby enhance the X-ray luminosity. Thermal conduction (and other sources of mass loading of the hot bubble, such as turbulent mixing) plays an important role in explaining the X-ray emission from galactic superbubbles because very little gas is expected to be in the X-ray emitting regime ($10^6-10^8$ K) in its absence (see the right panel of Fig. \ref{fig:dT_B_cond}).
\end{itemize}

Our simple one-dimensional simulations show that isolated supernova remnants, owing to large radiative losses, are much weaker feedback agents compared to superbubbles driven by coherently overlapping supernovae. However, detailed three-dimensional calculations, particularly with a realistic distribution of stars in a cluster, and magnetic fields and thermal conduction, are required in order to make quantitative comparisons with observations. This will be done in future.

\section*{Acknowledgements}
PS thanks Ramesh Narayan for useful discussions. PS is partly supported by DST-India grant no. Sr/S2/HEP-048/2012. YS is partly supported by RFBR (project code 12-02-00917).

\end{document}